\documentclass[12pt]{article}
\usepackage{epsfig}

\makeatletter \@addtoreset{equation}{section}

\def\href#1#2{#2}

\newcommand{\beq}{\begin{equation}}
\newcommand{\eeq}{\end{equation}}
\newcommand{\beqa}{\begin{eqnarray}}
\newcommand{\eeqa}{\end{eqnarray}}
\newcommand{\beqar}{\begin{eqnarray*}}
\newcommand{\eeqar}{\end{eqnarray*}}

\newcommand{\bC}{\bar{C}}
\newcommand{\bz}{\bar{z}}
\newcommand{\brho}{\bar{\rho}}

\newcommand{\ie}{{\it i.e.,}\ }
\newcommand{\labell}[1]{\label{#1}} 

\begin{document}

\thispagestyle{empty} \rightline{ \hfill
IPM/P-2004/039} \vspace*{4cm}

\begin{center}
{\bf \bf \Large Closed String S-matrix Elements in \\
Open String Field Theory}
\vspace*{2cm}

{Mohammad R. Garousi$^{}$\footnote{E-mail:garousi@ipm.ir} and
G.R. Maktabdaran$^{}$\footnote{E-mail:maktab@science1.um.ac.ir}}\\
\vspace*{0.2cm} {Department of Physics, Ferdowsi University,}
{ P.O. Box 1436, Mashhad, Iran}\\
\vspace*{0.1cm} {Institute for Studies in Theoretical
Physics and
Mathematics IPM} \\
{P.O. Box 19395-5531, Tehran, Iran}\\
\vspace*{0.4cm}

\vspace{2cm} ABSTRACT
\end{center}
We study  the  S-matrix elements of  the gauge invariant operators
corresponding to on-shell closed strings, in open string field
theory. In particular, we calculate the  tree level S-matrix
element of two ${\it arbitrary}$
 closed strings, and the
S-matrix element of one closed string and two  open strings. By
mapping the world-sheet of these amplitudes to the upper half
$z$-plane, and by evaluating explicitly  the correlators in the
ghost part, we show that these S-matrix elements are ${\it
exactly}$ identical to the corresponding disk level S-matrix
elements in perturbative string theory.

\vfill \setcounter{page}{0} \setcounter{footnote}{0}
\newpage

\section{Introduction}

The open string tachyon condensation has attracted much interest
recently. Regarding the recent development in string field theory
(see, for example, \cite{KO} and \cite{ABGKM} and their
references), it is fair  to say  that the Open String Field Theory
(OSFT) framework \cite{WITTEN} might provide a direct approach to
the study of  open string tachyon and could give striking
evidence for the tachyon condensation conjecture \cite{SEN}. This
framework is also appropriate for perturbative computation of
various open string S-matrix elements
\cite{giddings1,giddings2,Zwie}. There are several  methods for
evaluating the S-matrix elements \cite{wati,wati1}, however,
there is one which is suitable for analytic computation. This
method which is based on conformal mapping technique, has been
used to compute explicitly the Veneziano amplitude
\cite{giddings3,sloan,samuel1}. Furthermore, it has  been  shown
that any  loop amplitude can be evaluated with this method
\cite{samuel2,wati2}.

On the other hand, the most difficult part of the Sen's conjecture
\cite{SEN} is to understand how the closed strings emerge in the
true vacuum of tachyon potential. So it would be a natural
question to ask how the closed strings  appear in the OSFT. In
fact, it has been shown that the off-shell closed strings arise
in the OSFT because certain one-loop open string diagrams can be
cut in a manner that produces  closed string poles
\cite{FGST,wati3}. Unitarity then implies that they should also
appear as asymptotic states.
Closed string in the OSFT has been studied in several papers
including \cite{{TU},{ST1}, {STO}, {ZW}}.

In another attempt to incorporate the closed string into the OSFT,
some gauge invariant operators have been added into the OSFT
\cite{{HN}, {GRSW}}. These operators which are parameterized by
closed string vertex operators, add  on-shell closed strings into
the OSFT. It has
 been suggested in \cite{{HN},{GRSW}} that  the S-matrix elements of
these gauge invariant operators could be interpreted as the
scattering amplitude of their corresponding closed strings off a
D-brane. The S-matrix element of two such operators  has been
calculated and shown that the amplitude produces correctly the
pole structure of the  scattering amplitude of two closed strings
off a D-brane in the perturbative string theory \cite{MAMG}.
Similar analysis has been done for the S-matric element of one
gauge invariant operator and two open string vertex operators
\cite{MGGM}. In these papers \cite{MAMG,MGGM}, however, the
truncation method (up to level two) has been used to calculate
the amplitudes in the OSFT side.

The first exact   computation of  the S-matrix elements of the
gauge invariant operators which is
 based on the conformal
mapping technique, appears in \cite{TTSZ}. In this paper, the
S-matrix element of  two  tachyon operators  has been calculated
and shown to be in full agreement with the scattering amplitude
of two closed string tachyons off a D-brane in perturbative string
theory. The goal of the present paper is to use the conformal
mapping technique to evaluate  the S-matrix element of two ${\it
arbitrary}$ closed string operators, and the S-matrix element of
one ${\it arbitrary}$ closed string operator and two ${\it
arbitrary}$ open string vertex operators.

In the next section, we will briefly review   the OSFT and
 the gauge invariant operators. In section 3, we shall study the S-matrix element of two such operators.
  By mapping the world-sheet of the amplitude to the upper half $z$-plane  and by
evaluating explicitly the ghost correlators in the $z$-plane, we
shall show that the S-matrix element converts ${\it exactly}$ to
the  scattering amplitude  of two closed string states off a
D-brane in perturbative string theory.
 In section 4, we shall study the S-matrix element of one closed string operator and two open
strings vertex operators. Here also, by mapping the world-sheet
of the amplitude to the upper half $z$-plane and by performing the
ghost correlators, we shall show that the amplitude reproduces
${\it exactly}$  the amplitude describing the decay  of two open
strings to one closed string.

\section{Review of Open String Field Theory}

The OSFT is given by the following action \cite{WITTEN}:
 \beqa S_{open}(\Psi)&=&-\frac{1}{2\alpha'}\int \Psi *
Q\Psi-\frac{g_\circ}{3\alpha'} \int \Psi * \Psi *
\Psi\;\; ,\labell{SA} \eeqa
where $Q$ is the BRST charge with ghost number one, the open
string field $\Psi$ is a ghost number one state in the Hilbert
space of the first-quantized string theory, and $g_\circ$ is the
open string coupling. This action is invariant
under the gauge transformation $\delta\Psi=Q\Lambda+g_\circ(\Psi
* \Lambda-\Lambda * \Psi)$.

The gauge invariant operators in the OSFT  have been constructed
in \cite{HN,GRSW}. General form of these operators are given by
$g_c\int V \Psi$, where $g_c$ is the closed string coupling and
$V$ is an on-shell closed string vertex operator with conformal
dimension (0,0) and ghost number two. In order to be gauge
invariant, the closed string vertex operator has to be inserted
at the midpoint of open string.

This form of  gauge invariant operator can be understood from the
closed/open vertex studied in \cite{ZW}. It was shown that the
extended OSFT  with the action \beqa
S=-\frac{1}{2\alpha'}\left(\int \Psi * Q\Psi +{2g_o\over 3}\Psi
*\Psi *\Psi\right) +g_c\int V\Psi\;\; , \label{ACP} \eeqa where
$V$ is an on-shell closed string vertex defined at the midpoint of
the open string, would provide a theory which covers the full
moduli space of the scattering amplitudes of open and closed
string with a boundary \cite{ZW}. We note, however, that such
amplitudes  are actually the closed string scattering off a
D-brane. We should then be able to reproduce the closed string
scattering amplitudes in the framework of  open string field
theory.

To make sense  of the abstract form of the action, one may use
 the Fock space representation,
\beqa
S(|\Psi\rangle)&=&-\frac{1}{2\alpha'}\langle\Psi|Q|\Psi\rangle-
\frac{g_\circ}{3\alpha'}\langle
V_{123}||\Psi\rangle_{1}|\Psi\rangle_{2}|\Psi\rangle_{3}+
g_c\langle \mathcal{I}|V(\frac{\pi}{2})|\Psi\rangle \;\;
,\labell{FSA}\eeqa
where $|V_{123}\rangle$ is the 3-point vertex operator. We will carry out
the gauge-fixing by choosing the so-called Feynman-Siegel gauge $b_\circ
|\Psi\rangle=0$. In this gauge, the
BRST charge in the kinetic-energy term has a simple form,
\beqa Q&=&c_{0}L_{0}\;\; .\nonumber \labell{BRST} \eeqa
The  inverse of this operator is
 regarded as propagator, \ie
%
\beqa \frac{1}{Q} &=&  b_{0}\int_{0}^{\infty}dt e^{-t L_{0}} \;\;
,\labell{PR} \eeqa
where the modular parameter $t$ is the length of the propagating
open string.
\begin{figure}
\centerline{\psfig{file=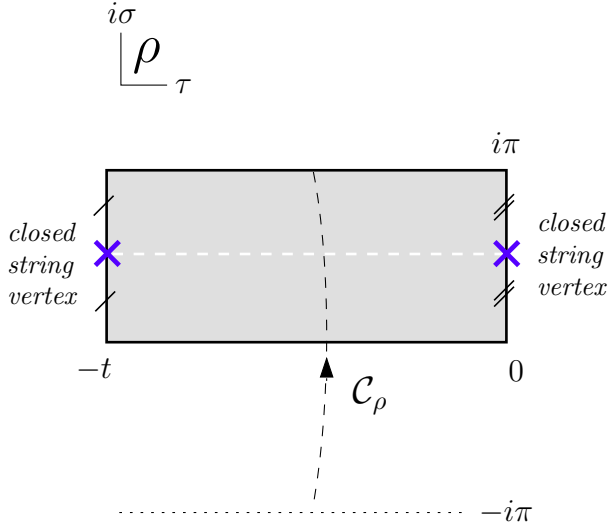}}\caption{Word-sheet description
of two closed string scattering amplitude. Contour $\mathcal
C_{\rho}$ is a path for ghost integration. \label{fig1}}
\end{figure}
%
%
%
In the last term in (\ref{FSA}), $|\mathcal{I}\rangle $ is the
identity string field which operates as  gluing  the two halves of
the string together. Hence, the open-closed vertex has the
following  geometrical interpretation: a closed string vertex
operator is inserted in the midpoint of the open string and then
the two halves of string are sewed together.

\section{Two closed string amplitudes}

Now using the Feynman rules in the previous section,
 we can calculate S-matrix elements of any closed or open
string. As has been  mentioned before, the S-matric element of
two closed string tachyon operators has been calculated
 in \cite{TTSZ}. The authors in \cite{TTSZ} have  used the conformal
mapping method  to calculate the amplitude exactly. Their result
is exactly the same as the scattering amplitude of two closed
string tachyons off a D-brane in perturbative string theory.
However, in this section we use a conformal map which is slightly
different from the one used in \cite{TTSZ}, and we show that the
S-matrix element of ${\it any}$ two closed strings is exactly
equal to the corresponding amplitude in perturbative bosonic
string theory.

The world-sheet of this scattering is shown in Fig.\ref{fig1}. It
is simply a finite length strip (open string propagator) where an
identity operation acts at each of the two points,  $\tau=0$ and
$\tau=-t$. The two closed string vertices should be inserted at
two singular midpoints. Using the propagator (\ref{PR}) one may
write the  amplitude as \footnote{The coupling constants
$g_\circ$ and $g_c$  appear as multiplication factors in
amplitudes. However, for simplicity we will ignore them
hereafter.}:
\beqa A_{O_c O_c}&=& \int_{0}^{\infty}dt \langle
V_{12}|b_{0}^{}e^{-t L_{0}^{(2)}}
|O_c\rangle_{1}|O_c\rangle_{2} \;\; ,\labell{CCAMP1} \eeqa
where $|V_{12}\rangle$ is the 2-point vertex operator and
$|O_c\rangle \equiv V_{O_c}(\frac{i\pi}{2})|\mathcal{I}\rangle$
defines  a closed string state. The above amplitude can be written
in terms of CFT correlation functions in the $\rho$-plane as
\beqa
A_{O_cO_c}&=&\int_0^{\infty}dt\int_{C_{\rho}}\frac{d\rho}{2\pi i}
\langle
b_{\rho\rho}(\rho)V_{1O_c}(\frac{i\pi}{2})V_{2O_c}(\frac{i\pi}{2}-t)\rangle_{\rho}
\;\; ,\labell{aconf}\eeqa
where we have used $b_0=\int_{\mathcal{C}_{\rho}}\frac{d \rho}{2
\pi i} b_{\rho\rho}(\rho)$ and the contour $\mathcal{C}_{\rho}$ is
from $-i\pi$ to $i\pi$. In writing this relation we have used the
doubling trick
$b_{\bar{\rho}\bar{\rho}}(\bar{\rho})=b_{\rho\rho}(\bar{\rho})$ as
well. The closed string vertex operator is
\beqa V_{O_c}(\rho)&=&C(\rho)\bC(\brho)O_c(\rho,\brho)\;\;
,\labell{cvertx}\eeqa
where $O_c$ is the matter part of the vertex operator. Using the
fact that world-sheet propagators  have simple forms in the upper
half $z$-plane, one usually maps the world-sheet Fig.\ref{fig1} to
the upper half $z$-plane. A map which does this has been given in
\cite{TTSZ} to calculate the amplitude for two tachyons. However,
we will map this world-sheet to the $z$-plane in a slightly
different way which makes us  able to conclude that the amplitude
of any two closed strings is the same as the corresponding
amplitude in perturbative string theory.
\begin{figure}
\centerline{\psfig{file=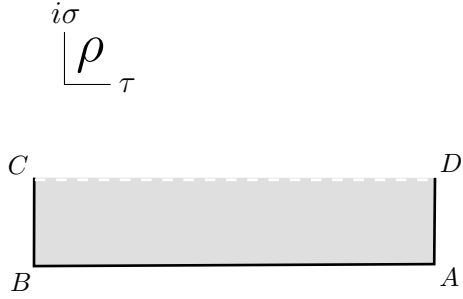}}\caption{Half part of the
Riemann surface depicted in Fig. \ref{fig1}.\label{fig2}}
\end{figure}
%


\subsection{Conformal mapping }

By cutting the Riemann world-sheet along the line joining the two
singular pints, we can divide it into two similar rectangles, with
no singularity therein. We call them $ABCD$ and $A'B'C'D'$. The
rectangle $ABCD$ is shown in Fig.\ref{fig2}. Gluing  segments
$BC$ to $B'C'$, $CD$ to $C'D'$, and $DA$ to $D'A'$,  one will
restore the original world-sheet in Fig.\ref{fig1}. The basic
idea of mapping the world-sheet to $z$-plane is to find a map
which brings the upper right half $z$-plane to  the rectangle
$ABCD$ and the upper left half $z$-plane to the rectangle
$A'B'C'D'$ in such a way that  the image of segments $BCDA$ and
$B'C'D'A'$ each separately covers the imaginary axis completely.
Since  the imaginary axis is common between  upper right half
$z$-plane and upper left half $z$-plane, this axis  shows the
gluing part of the original world-sheet in Fig.\ref{fig1}.

\begin{figure}
\centerline{\psfig{file=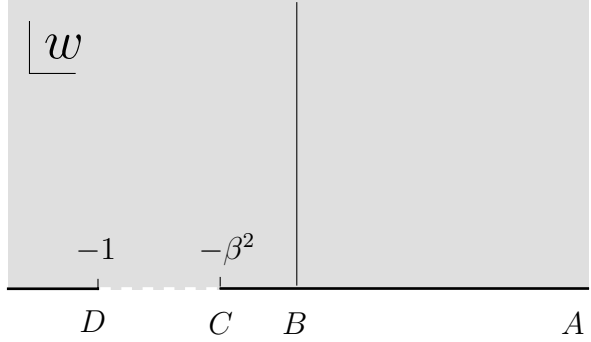}}\caption{Image of the rectangle
depicted in Fig. \ref{fig2} on the $w$-plane. \label{fig3}}
\end{figure}

Now we try to find the map in two steps. Firstly, the
Schwarz-Christoffel method tells us  how to find the map
$\rho(w)$, which brings the upper half $w$-plane into the
interior of any polygon, so that the image of real axis covers
the circumference of the polygon once. Derivative of this map is
$\frac{d\rho}{dw}= \frac{N}{2}\prod_{i=1}^{i=m}(w-w_i)^{a_i}$
where $w_i$'s are images of the vertices of the polygon, and $N$
is a constant. Edges of the vertex $i$ rotate  through the angle
$a_i\pi$ which is positive (negative) for counterclockwise
(clockwise) direction. For the rectangle $ABCD$ in Fig.\ref{fig2}
this becomes
\beqa \frac{d\rho}{d
w}&=&\frac{N}{2}\frac{1}{\sqrt{w}\sqrt{w+\beta^{2}}
\sqrt{w+1}}\;\; ,\labell{CCSCTW} \eeqa
in which we have assumed that points ${\infty}$, ${0}$,
${-\beta^2}$, and $-1$ on the $w$-plane are images of vertices
${A}$, ${B}$, ${C}$, and ${D}$, respectively (see Fig.\ref{fig3}).
Note that the $SL(2,R)$ symmetry of the upper half $w$-plane
allows us to set three points fixed. We chose them to be the
images of ${A}$ and ${B}$, and $D$\footnote{If one assumes that
points $\infty$, $0$, $-\beta^2$, and $-1/\beta^2$ on the
$w$-plane are images of vertices A, B, C, and D, respectively,
then one would recover the map used in \cite{TTSZ}. In that case,
$SL(2,R)$ symmetry of the upper half $w$-plane is fixed by fixing
the images of A and B, and fixing the relation between the two
remaining vertices.}. Using the fact that all points at infinity
are coincident, the above transformation maps segment $BCDA$ to
the negative real axis of the $w$-plane. The square roots in the
denominator of (\ref{CCSCTW}) come from the fact that all edges
of the rectangle rotate  $\pi/2$ clockwise. We have also absorbed
the constant $1/\sqrt{w-\infty}$ in $N$. We shall find the
constant $N$ shortly.

\begin{figure}
\centerline{\psfig{file=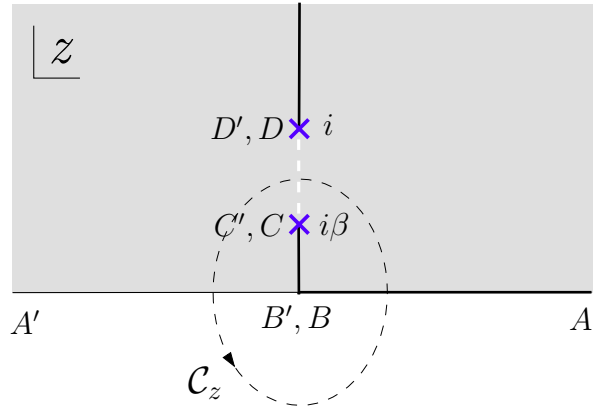}}\caption{Upper half $z$-plane
description of the world-sheet in Fig. \ref{fig1}. The conformal
map $(\ref{CCSCT})$ transforms the $z$-plane to the Riemann
surface depicted in Fig. \ref{fig1}. Contour $\mathcal C_z$ is a
path for ghost integration. \label{fig4}}
\end{figure}

The second step is to  find a transformation which maps the upper half $w$-plane
to  the  right  hand side of the upper half $z$-plane. This map  is
simply $w=z^{2}$. Combining the two transformations $\rho(w)$ and
$w(z)$ will bring the  right  hand side of the upper half $z$-plane to
the rectangle $ABCD$ such that the imaginary axis of the $z$-plane
maps to the segment $BCDA$, as required. The map is then,
\beqa \rho(z)&=&\int \left(\frac{d\rho}{dw}\right)\left(\frac{dw}{dz}\right)dz\nonumber\\
&=& N\int^{z}_{z_{0}}\frac{dz}{\sqrt{z^{2}+\beta^{2}}
\sqrt{z^{2}+1}}\;\; .\labell{CCSCT} \eeqa
As shown in the Fig.\ref{fig4}, images of the vertices on the
$z$-plane are ${\infty}$, ${0}$, ${i\beta}$, and $i$. In above
equation $z_{0}$ is a constant in which we are not interested. The
same map will bring the upper left hand side of $z$-plane to the
rectangle $A'B'C'D'$ in which the imaginary axis is mapped to
segment $B'C'D'A'$. Therefore, one recovers the original
world-sheet in Fig.\ref{fig1} in which the segments $BCDA$ and
$B'C'D'A'$ are glued together.

To find the constant $N$, we use the fact that the length of
segment $BC$ must be equal to $\pi/2$. This gives
\beqa \rho(i\beta)-\rho(0)&=& N
\int^{i\beta}_{0}\frac{dz}{\sqrt{z^{2}+\beta^{2}}
\sqrt{z^{2}+1}}= \frac{i\pi}{2}\;\; .\nonumber \labell{N1} \eeqa
Using the change of variable as $u=z/(i\beta)$, this integral
becomes of the form of a complete elliptic integral of the first
kind
\footnote{$K(m)=\int_0^1du\frac{1}{\sqrt{1-u^2}\sqrt{1-mu^2}}$}\cite{HandBook}.
Hence,  one finds,
\beqa N&=&\frac{\pi}{2 K(m)} \;\; ,\labell{CCNORMA} \eeqa
where $m=\beta^{2}$.

The idea of conformal mapping is to map the positions of all
operators in (\ref{aconf}) to the $z$-plane, and also change the
modular parameter $t$ to the modular parameter $\beta$ which is
the position of one of  the vertex operators. To find this
relation between $\beta$ and $t$, we use the fact that the length
of segment $CD$ is equal to $t$, that is,
\beqa \rho(i)-\rho(i\beta)&=& N
\int^{i}_{i\beta}\frac{dz}{\sqrt{z^{2}+\beta^{2}}
\sqrt{z^{2}+1}}= t \;\; .\nonumber \labell{T1} \eeqa
Using (\ref{CCNORMA}) and changing the variable as
$u^{2}=(1+z^{2})/(1-\beta^{2})$, one finds an  explicit relation,
\beqa \frac{\pi K(m')}{2 K(m)}&=&t \;\; ,\labell{T2} \eeqa
in which $m'=1-m$. It is easy to see that $\beta$ goes to zero
when $t\rightarrow \infty$, and goes to one as $t\rightarrow 0$.
Using the derivative formula
\beqa dK(m)/dm=[E(m)-m'K(m)]/2mm'\labell{derv1}\;\; ,\eeqa
where $E$ is the complete elliptic integral of the second kind, and
the Legendre's relation,
\beqa E(m')K(m)-K(m')K(m)+E(m)K(m')&=&\pi/2 \;\; ,\labell{LEGEN}
\eeqa
one can easily differentiate (\ref{T2}) to get,
\beqa dt&=-&\frac{(\pi/2)^2}{K^2(m)\beta(1-\beta^{2})}d\beta \;\;
.\labell{CCDMEAS} \eeqa
Note that the non trivial function $K(m)$ appears in the Jacobian of the transformation.
However, as we shall see shortly, this function does not appear in the final amplitude.
%

Finally,  using the fact that the anti-ghost field $b_{\rho\rho}$
has conformal weight 2,
one can write the following transformation:
\beqa
\int_{\mathcal{C}_{\rho}}\frac{d \rho}{2 \pi i}
b_{\rho\rho}(\rho)=\oint_{\mathcal{C}_{z}}\frac{d z}{2 \pi
i}\left(\frac{d z}{d\rho}\right) b_{zz}(z)\;\; .
\labell{bzero} \eeqa

Now, using equations (\ref{bzero}) and (\ref{CCDMEAS}), and using
the fact that vertex 1 (2) at point $i\pi/2$ ( $i\pi/2-t$) in the
$\rho$-plane is mapped to $i$ ( $i\beta$) in the $z$-plane, one
can write the scattering amplitude (\ref{aconf}) in the $z$-plane
as
\beqa A_{O_cO_c}&=&\int_0^1 d\beta\langle
O_{1c}(i,-i)O_{2c}(i\beta,-i\beta)\rangle_z
\labell{aconfz}\\
&\times&\frac{(\pi/2)^2}{\beta(1-\beta^2)K^2(m)}\oint_{\mathcal{C}_{z}}
\frac{dz}{2\pi i}\left( \frac{dz}{d\rho}\right)\langle
C(i)\bC(-i)b(z)C(i\beta)\bC(-i\beta)\rangle_z \;\; , \nonumber\eeqa
where $b(z)=b_{zz}(z)$.

Using the doubling trick $\bC(\bz)=C(\bz)$, the two-point
function $\langle b(z)C(z') \rangle=1/(z-z')$, and the relation
$\langle C(z_1)C(z_2) C(z_3)
\rangle=(z_1-z_2)(z_2-z_3)(z_1-z_3)$, one finds
\beqa \langle C(i)\bC(-i)b(z)C(i\beta)\bC(-i\beta)\rangle_z&=
&-\frac{4\beta(1-\beta^2)^2}{(z^2+\beta^2)(z^2+1)} \;\; .\eeqa
Inserting this in the integral (\ref{aconfz}), and using equations
(\ref{CCSCT}) and (\ref{CCNORMA}), one can perform the
z-integration in (\ref{aconfz}). The result is
\beqa A_{O_cO_c}&=-4&\int_0^1 d\beta\, (1-\beta^2) \langle
O_{1c}(i,-i)O_{2c}(i\beta,-i\beta)\rangle_z \;\; ,\labell{aconff}
\eeqa
where we have used the formula
\beqa \frac{1}{2 \pi i}\oint_{\mathcal{C}_{z}}\frac{d z
}{\sqrt{(z^{2}+\beta^{2})(z^{2}+1)}}&=&\frac{2}{\pi}\int_0^1\frac{du}{\sqrt{(1-u^2)(1-\beta^2
u^2)}}\nonumber\\
&= &\frac{2 K(m)}{\pi} \;\; . \labell{KINT} \eeqa
In finding the above formula, we have used the fact that the
contour $\mathcal C_z$ includes the two singular points
$z=i\beta$ and $z=-i\beta$. As we have mentioned before, the
function $K(m)$ does not appear in the final amplitude
(\ref{aconff}).

On the other hand,   the S-matrix element describing the
scattering amplitude of two arbitrary closed string states off a
D-brane in perturbative string theory is given by the following
correlation function in the $z$-plane:
\beqa A_{O_cO_c}\sim \int d^2z_1d^2z_2\langle
O_{1c}(z_1,\bz_1)O_{2c}(z_2,\bz_2)\rangle_z \;\; .\labell{samp}
\eeqa
The integrand has SL(2,R) symmetry. Gauging this symmetry by
fixing $z_1=i$ and $z_2=i\beta$, one finds \cite{GM}
\beqa \int d^2z_1d^2z_2\longrightarrow
\int_{0}^1d\beta\,(1-\beta^2) \;\; .\eeqa
This makes the amplitude (\ref{samp}) to be exactly like the
amplitude in (\ref{aconff}). This completes  our illustration of
the equality between
 the tree level  S-matrix elements involving two arbitrary
closed string states in the OSFT and the corresponding disk level
S-matrix elements in perturbative string theory.

\section{One closed and two open strings amplitudes}
\begin{figure}
\centerline{\psfig{file=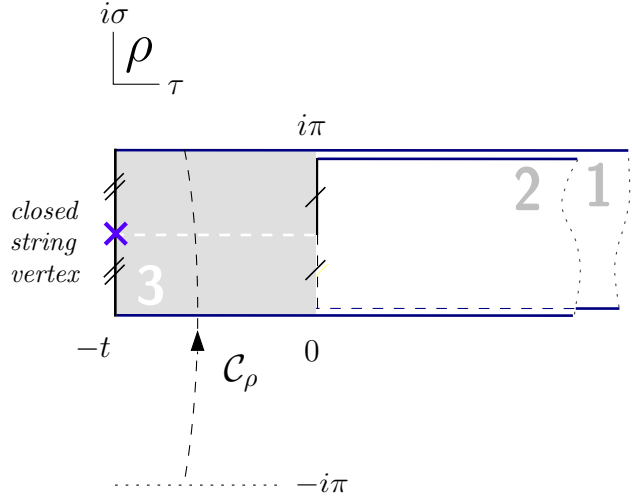}}\caption{Word-sheet description
of one closed and two open string scattering amplitude. Contour
$\mathcal C_{\rho}$ is a path for ghost integration. \label{fig5}}
\end{figure}

The more involved amplitude is the S-matrix element of  one
closed and two open strings. World-sheet description for this
process is depicted in Fig.\ref{fig5}. Two on-shell open strings
(semi-infinite strip) and one off-shell open string (finite
length strip) are gluing together at $\tau=0$, according to the
Witten's way of joining strings \cite{WITTEN}.
At time $\tau=-t$, where $t$ is between $0$ and $\infty$, a closed
string vertex operator is inserted on the singular midpoint
generated by the identity operator acting  on  string 3. Using the
propagator (\ref{PR}), one may write the amplitude as:
\beqa A_{O_oO_oO_c}&=& \int_{0}^{\infty}dt \langle
V_{123}|b_{0}^{}e^{-t L_{0}^{(3)}}
|O_o\rangle_{1}|O_o\rangle_{2}|O_c\rangle_{3} \;\; ,\labell{AMP1}
\eeqa
where $|V_{123}\rangle$ is the 3-point vertex operator, and
$|O_o\rangle$ is an open string state. The above amplitude can be
written in terms of CFT correlation functions in the $\rho$-plane
as
\beqa
A_{O_oO_oO_c}&=&\int_0^{\infty}dt\int_{C_{\rho}}\frac{d\rho}{2\pi
i} \langle b_{\rho\rho}(\rho)V_{1O_o}(\infty)V_{2O_o}(\infty)
V_{3O_c}(\frac{i\pi}{2}-t)\rangle_{\rho} \;\;
.\labell{aconf1}\eeqa
The closed string vertex operator is given in (\ref{cvertx}), and
the open string vertex operator is
\beqa V_{O_o}(\rho)&=&C(\rho)O_o(\rho)\;\; ,\eeqa
where $O_o$ is the matter part of the open string vertex
operator. Searching for a suitable transformation that maps  the
world-sheet to the upper half $z$-plane, is a crucial step for
performing the above CFT computations. We will follow the same
strategy as in the previous section.

\subsection{Conformal Mapping }
\begin{figure}
\centerline{\psfig{file=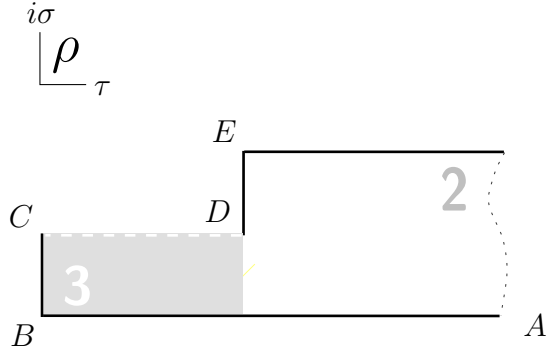}}\caption{Half part of the
Riemann surface depicted in Fig. \ref{fig5}.\label{fig6}}
\end{figure}

\begin{figure}
\centerline{\psfig{file=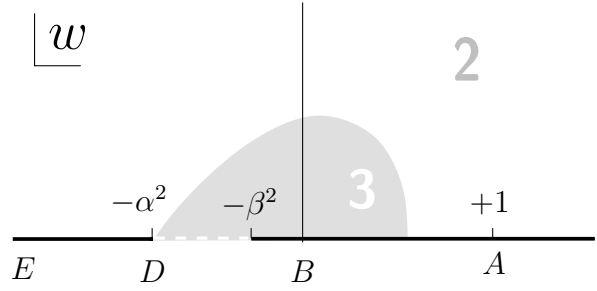}}\caption{Image of the pentagon
depicted in Fig. \ref{fig6} on the $w$-plane. \label{fig7}}
\end{figure}

As in the two closed strings  case, here also we face  a Riemann
surface with two singular points. By cutting the world-sheet
along the line joining the two singular pints, we can divide it
into two similar pentagons, with no singularity therein. We call
them $ABCDE$ and $A'B'C'D'E'$. One of these degenerated pentagons
is depicted in Fig. \ref{fig6}. Vertex $A$ goes to plus infinity.
Joining the two pentagons along the segments $BCDE$ and
$B'C'D'E'$ will restore the original world sheet. We should find
a map $\rho(z)$ which brings the upper right half $z$-plane to
one pentagon, so that the image of segment $BCDE$ covers the
imaginary axis completely. Again we do it in two steps. By the
Schwarz-Christoffel method we can find the map $\rho(w)$, which
brings the upper half $w$-plane into the interior of the
pentagon. Regarding all exterior angles, one can write
\beqa \frac{d\rho}{d
w}&=&\frac{N}{2}\frac{\sqrt{w+\alpha^2}}{\sqrt{w}\sqrt{w+\beta^{2}}(w-
1)}\;\; ,\labell{SCTW} \eeqa
in which we assume that points $1$, ${0}$, ${-\beta^2}$,
${-\alpha^2}$ and ${-\infty}$ on the $w$-plane are images of vertices
${A}$, ${B}$, ${C}$, ${D}$ and ${E}$, respectively ( see Fig.\ref{fig7}).

\begin{figure}
\centerline{\psfig{file=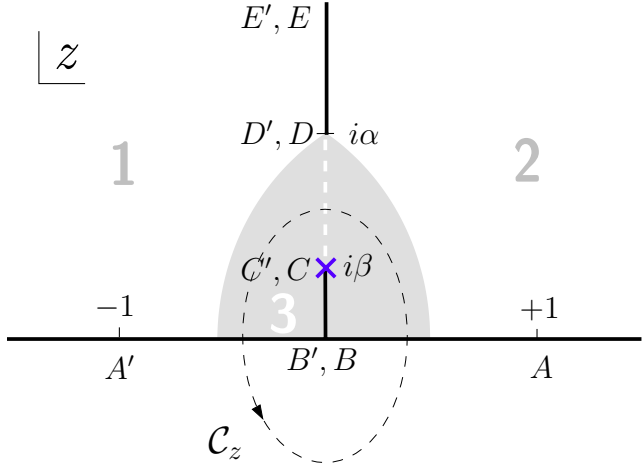}}\caption{Upper half $z$-plane
description of the world-sheet in Fig. \ref{fig5}. The conformal
map $(\ref{SCT})$ transforms the $z$-plane to the Riemann surface
depicted in Fig. \ref{fig5}. Contour $\mathcal{C}_{z}$ is a path
for ghost integration. \label{fig8}}
\end{figure}

Combining $\rho(w)$ with transformation $w=z^{2}$, one finds
$\rho(z)$:
\beqa \rho(z)&=& N
\int^{z}_{z_{0}}dz\frac{\sqrt{z^{2}+\alpha^2}}{\sqrt{z^{2}+\beta^{2}}(z^{2}-
1)}\;\; .\labell{SCT} \eeqa
As shown in the Fig.\ref{fig8}, images of the pentagon vertices on
the $z$-plane are $1$, ${0}$, ${i\beta}$, ${i\alpha}$ and
${+\infty}$. To find the constant $N$, one may consider the fact
that the weight of strip 2 at $A$ must be equal to $\pi$. This
gives
\beqa \int dz(\frac{d \rho}{dz})&=& i\pi \;\; ,\nonumber
\labell{CI} \eeqa
where the integration is taken along a clockwise semi-contour
around the point $A$ in the upper half $z$-plane. This indicates
that  $d \rho /dz $ must have a simple pole at $z=1$,
\beqa \frac{d \rho}{dz}|_{z=1}&=&\frac{-1}{z-1} \;\; .\nonumber
\labell{POL} \eeqa
This requirement will fix  the factor $N$ to be
\beqa N&=&-2\frac{\sqrt{1+\beta^2}}{\sqrt{1+\alpha^2}} \;\;
.\nonumber \labell{NORMA} \eeqa

For the special case that $\alpha=\beta$, one has a semi-infinite
strip which maps to the right hand side of the upper half
$z$-plane by $\rho(z)= -\ln(1-z)+\ln(1+z)$. This fixes $z_0$ in
(\ref{SCT}) to be  $z_0=0$. This transformation also maps the mid
point of the semi-infinite strip to $z=i$.

The parameters $\alpha$ and $\beta$ are both functions of $t$.
Hence there should be a relation between these two parameters. To
find this relation, one may consider  the condition that the
length of segment $BC$ must be $\pi /2$. This leads to the
following constraint equation between $\alpha$ and $\beta$:
\beqa f(\alpha,\beta)&=& \frac{i\pi}{2}\;\; ,\labell{CON1} \eeqa
where $f(\alpha,\beta)\equiv\rho(i \beta)-\rho(0)$.

Making use of  (\ref{SCT}), one can write $f(\alpha,\beta)$ as
\beqa f(\alpha,\beta)&=& N \int^{i \beta
}_{0}dz\frac{\sqrt{z^{2}+\alpha^2}}{\sqrt{z^{2}+\beta^{2}}(z^{2}-
1)}\;\; ,\nonumber \labell{F1} \eeqa
which can be rewritten in terms of complete elliptic integrals by
using the change of variable $u=z/i \beta$,
\beqa f(\alpha,\beta)=i N\alpha \left[\frac{1}{\alpha^2}\int^{1
}_{0}\frac{du}{\sqrt{1-u^{2}}\sqrt{1-m
u^{2}}}-(1+\frac{1}{\alpha^{2}})\int^{1
}_{0}\frac{du}{\sqrt{1-u^{2}}\sqrt{1-m u^{2}}(1-n u^{2})}
\right],\nonumber \labell{F2} \eeqa
where $m=\beta^{2}/\alpha^2$ and $n=-\beta^{2}$. The first and
second integrals are the standard form of complete elliptic
integrals of  first and third kind, respectively. Hence,
\beqa f(\alpha,\beta)&=&i N\alpha
\left[\frac{1}{\alpha^2}K(m)-(1+\frac{1}{\alpha^{2}})\Pi(n|m)
\right]\;\; .\nonumber \labell{F3} \eeqa
Using the Jacobi formula\footnote{
$\Pi(n|m)=K(m)-\frac{tan\phi}{\sqrt{1-n}}
\left[E(m)F(\phi|m)-K(m)E(\phi|m)\right]$ where
$\phi=sin^{-1}(\sqrt{\frac{n}{m}})$} it is possible to write
$\Pi(n|m)$ in terms of elliptic integrals of  first and second
kind \cite{HandBook}. After some algebra, one gets,
\beqa f(\alpha,\beta)&=& 2
K(m)\left[i\alpha\sqrt{\frac{1+\beta^2}{1+\alpha^2}}-Z(
\phi|m)\right]\;\; ,\nonumber \labell{F4}\eeqa
where $Z$ is the Jacobi Zeta
function\footnote{$Z(\phi|m)=E(\phi|m)-\frac{E(m)}{K(m)}F(\phi|m)
$} \cite{HandBook} and $\sin{\phi}=i\alpha$.
%
%

Another constraint equation which sets relation between $\alpha$,
$\beta$ and $t$, comes from the condition that the length of
segment $CD$ must be equal to $t$, that is,
\beqa g(\alpha,\beta)&=& t \;\; ,\labell{CON2} \eeqa
where $g(\alpha,\beta)\equiv\rho(i\alpha)-\rho(i \beta)$.

Using (\ref{SCT}), one can write $g(\alpha,\beta)$ as
\beqa g(\alpha,\beta)&=& N \int^{i\alpha}_{i
\beta}dz\frac{\sqrt{z^{2}+\alpha^2}}{\sqrt{z^{2}+\beta^{2}}(z^{2}-
1)} \;\; .\nonumber \labell{G1} \eeqa
Again this can be rewritten in terms of complete elliptic
integrals using the change of variable
$u^{2}=(\alpha^2+z^{2})/(\alpha^2-\beta^{2})$,
\beqa g(\alpha,\beta)&=& \frac{N}{\alpha} \left[\int^{1
}_{0}\frac{du}{\sqrt{1-u^{2}}\sqrt{1-m' u^{2}}}-\int^{1
}_{0}\frac{du}{\sqrt{1-u^{2}}\sqrt{1-m' u^{2}}(1-n' u^{2})}
\right]\;\;  \nonumber\\
&=& \frac{N}{\alpha} \left[K(m')-\Pi(n'|m')\right]\;\; \nonumber\\
&=& 2K(m')Z(\phi'|m')\;\; ,\nonumber \labell{G2} \eeqa
where $m'=1-m$, $n'=(\alpha^2-\beta^2)/(1+\alpha^2)$,
and $\sin{\phi'}=\alpha/\sqrt{1+\alpha^2}$.
From above relations, one can find the boundary value for $\beta$.
This boundary is $\beta\rightarrow 0$ as $t\rightarrow \infty$,
and $\beta=\alpha$ at $t=0$. However, as we have shown  before,
the point $\alpha=\beta$ which is the midpoint of the
semi-infinite strip is mapped to $z=i$. Hence, at $t=0$,
$\beta\rightarrow 1$.


Making use of  the two  constraint equations (\ref{CON1}) and
(\ref{CON2}), one can write the measure factor $dt$ in terms of
$d\beta$, that is,
\beqa dt &=& \left(\partial_{\beta}g -\frac{\partial_{\beta}f}{\partial_{\alpha}f}
\partial_{\alpha}g\right)d\beta
\;\; .\labell{DT1} \eeqa
Using the derivative formulas (\ref{derv1}) and
\beqa
dZ(\phi|m)&=&\left\{\frac{1}{2m}\left(1-\frac{E(m)}{m'K(m)}\right)Z(\phi|m)+
\frac{E(m)\sin(2\phi)}{4m'K(m)\sqrt{1-m\sin^2(\phi)}}\right\}dm\nonumber\\
&&+\left\{\sqrt{1-m\sin^2(\phi)}-\frac{E(m)}{K(m)\sqrt{1-m\sin^2(\phi)}}\right\}d\phi
\;\; ,\eeqa
one can   show, after some algebra,  that
\beqa \partial_{\alpha}g
&=&\frac{2}{1+\alpha^{2}}\sqrt{\frac{1+\beta^2}{1+\alpha^2}}K(m')
\;\; , \nonumber\\
\partial_{\beta}g
&=&\frac{-2(\alpha/\beta)}{\sqrt{(1+\alpha^{2})(1+\beta^{2})}}E(m')
\;\; ,\labell{DT2}\\
\partial_{\alpha}f&=&\frac{2i}{1+\alpha^{2}}\sqrt{\frac{1+\beta^2}{1+\alpha^2}}K(m)
\;\; ,\nonumber\\
\partial_{\beta}f
&=&\frac{2i(\alpha/\beta)}{\sqrt{(1+\alpha^{2})(1+\beta^{2})}}\left(E(m)-K(m)\right)
\;\; .\nonumber\eeqa
Inserting (\ref{DT2}) into (\ref{DT1}) and using the identity
(\ref{LEGEN}), one finds the following relation between $dt$ and
$d\beta$:
\beqa dt =-
\frac{\pi(\alpha/\beta)}{K(m)\sqrt{(1+\alpha^{2})(1+\beta^{2})}}d\beta
  \;\; .\labell{A11} \eeqa

Now making use of   equations (\ref{bzero}) and (\ref{A11}), and
use the fact that vertices 1, 2, and 3 are mapped  to points -1,
1, and $i\beta$, respectively, one can write
 the scattering amplitude (\ref{aconf1}) in the $z$-plane
as
\beqa A_{O_oO_oO_c}&=&\int_0^1 d\beta\langle
O_{1o}(-1)O_{2o}(1)O_{3c}(i\beta,-i\beta)\rangle_z
\labell{aconff2}\\
&\times&\frac{\pi(\alpha/\beta)}{K(m)\sqrt{(1+\alpha^2)(1+\beta^2)}}
\oint_{\mathcal{C}_{z}} \frac{dz}{2\pi i}\left(
\frac{dz}{d\rho}\right)\langle
C(-1)C(1)b(z)C(i\beta)\bC(-i\beta)\rangle_z \;\; .\nonumber\eeqa
The correlation in the ghost part can be evaluated. One finds,
\beqa \langle C(-1)C(1)b(z)C(i\beta)\bC(-i\beta)\rangle_z&=
&\frac{4i\beta(1+\beta^2)^2}{(z^2+\beta^2)(1-z^2)} \;\; .\eeqa
Inserting this into (\ref{aconff2}), using the map (\ref{SCT}) to
evaluate $\frac{dz}{d\rho}$, and   performing the z-integration
using (\ref{KINT}), one finds finally
\beqa A_{O_oO_oO_c}&=4i&\int_0^1 d\beta\, (1+\beta^2) \langle
O_{1o}(-1)O_{2o}(1)O_{3c}(i\beta,-i\beta)\rangle_z \;\;
.\labell{aconff1} \eeqa
It is interesting to note that here also the non trivial
function $K(m)$ which appears in the measure $dt$ (\ref{A11}), is
canceled by the same function that results from integration over
the ghost correlators. Similar cancellation happens  for the
S-matrix element of four tachyons too \cite{giddings3}.

In the perturbative  string theory, on the other side, the
S-matrix element describing the decay amplitude of two arbitrary
open string states on a D-brane to one arbitrary closed string
state is given by the following correlation function in the
$z$-plane:
\beqa A_{O_oO_oO_c}\sim \int dx_1dx_2d^2z_3\langle
O_{1o}(x_1)O_{2o}(x_2) O_{3c}(z_3,\bz_3)\rangle_z \;\;
.\labell{samp1} \eeqa
The integrand has again SL(2,R) symmetry. Gauging this symmetry
by fixing $x_1=-1$ $x_2=1$, and $z_3=i\beta$, one finds \cite{HK},
\beqa \int dx_1dx_2d^2z_3\longrightarrow
\int_{0}^1d\beta\,(1+\beta^2) \;\; .\eeqa
This makes the amplitude (\ref{samp1}) to be exactly like the
amplitude in (\ref{aconff1}). This completes  our illustration of
the equality between
 the tree level  S-matrix elements involving one arbitrary closed string and two arbitrary
 open
strings in the OSFT and the corresponding disk level S-matrix
elements in perturbative string theory.

It has been shown  in \cite{giddings3} that the  S-matrix element
of four tachyons in the OSFT is identical to the corresponding
amplitude in perturbative string theory. Making use of the
strategy utilized in the present paper, one can extend that
equality to the S-matrix element of four ${\it arbitrary}$ open
string states.

Finally, one may try to extend the S-matrix element of one closed and two open strings
 to the off-shell
physics using the method in \cite{sloan}. The important point is
that only open string states can be extended to the off-shell. It
is of interest to investigate  such an  off-shell extension. It
would also be interesting  to study
 higher point functions involving the closed string operators.

{\bf Acknowledgments}:
M.R.G would  like to thank
ICTP for hospitality during the completion of this work.
 G.R.M was supported by Birjand
university. We would like to thank A.E. Mosaffa for critical
reading of the  manuscript.


\begin{thebibliography}{99}

\bibitem{KO}
K. Ohmori, ``A Review on Tachyon Condensation in Open String
Field Theories'', hep-th/0102085.

\bibitem{ABGKM}
I.Ya. Aref'eva, D.M. Belov, A.A. Giryavets, A.S. Koshelev and
P.B. Medvedev, ``Non-commutative Field Theories and (Super)String
Field Theories'', hep-th/0111208.


\bibitem{WITTEN}
E. Witten, ``Non-commutative geometry and string field theory'',
Nucl. Phys. B {\bf 268} (1986) 253.
\bibitem{SEN}
A. Sen, `` Descent Relations Among Bosonic D-branes'',   Int. J. Mod. Phys. A
{\bf  14}  (1999) 4061 [arXiv:hep-th/9902105]; `` Universality of the
Tachyon Potential'',
 JHEP {\bf 9912} (1999) 027
[arXiv:hep-th/9911116].

\bibitem{giddings1}
S.B. Giddings and E. J. Martinec, ``Conformal Geometry and
String Field Theory'', Nucl. Phys. B{\bf 278} (1986) 91.
\bibitem{giddings2}
S.B. Giddings, E.J. Martinec and E. Witten, ``Modular Invariance
In String Field Theory'', Phys. Lett. B {\bf 176} (1986) 362.
\bibitem{Zwie}
B. Zwiebach, ``A Proof That Witten's Open String Theory Gives A Single Cover Of Moduli
Space'', Commun. Math. Phys. {\bf 142} (1991) 193.
\bibitem{wati}
W. Taylor, ``Perturbative diagrams in string field theory'',
arXiv:hep-th/0207132.
\bibitem{wati1}
W. Taylor and B. Zwiebach, ``D-Branes, Tachyons, and String Field
Theory'', arXiv:hep-th/0311017.


\bibitem{giddings3}
S.B. Giddings, ``The Veneziano amplitude from interacting string
field theory ``, Nucl. Phys. B {\bf 278} (1986) 242.
\bibitem{sloan}
J.H. Sloan, ``The scattering amplitude for four off-shell
tachyons from functional integrals '', Nucl. Phys. B {\bf 302}
(1988) 349.
\bibitem{samuel1}
S. Samuel, ``Covariant off-shell string amplitudes '', Nucl. Phys. B
{\bf 308} (1988) 285.
\bibitem{samuel2}
S. Samuel, ``Solving The Open Bosonic String In Perturbation
Theory'', Nucl. Phys. B {\bf 341} (1990) 513.
\bibitem{wati2}
W. Taylor, ``Perturbative computations in string field theory'',
arXiv:hep-th/0404102.
\bibitem{FGST}
D.Z. Freedman, S.B. Giddings, J.A. Shapiro and C.B. Thorn,
``The Nonplanar One Loop Amplitude In Witten's String Field
Theory'', Nucl. Phys. B {\bf 298} (1988) 253.

\bibitem{wati3}
W. Taylor , ``Tadpoles and Closed String Backgrounds in Open
String Field Theory'', JHEP {\bf 0307} (2003) 059
[arXiv:hep-th/0304259].

\bibitem{TU}
H. Terao and S. Uehara, ``On the dilaton vertex in the Covariant
Formulation of String'', {\em Phys. Lett. } {\bf B188} (1987) 198.

\bibitem{ST1}
J. A. Shapiro  and C. B. Thorn, ``BRST invariant Transition
between Closed and Open String'', {\em Phys. Rev.} {\bf D36}
(1987) 432, and ``Closed String-Open String Transition and
Witten's String Field Theory'', {\em Phys. Lett.} {\bf B194}
(1987) 43.

\bibitem{STO}
A. Strominger, ``Closed String in Open String Field Theory'',
 {\em Phys. Rev. Lett.} {\bf 16} (1987) 629.

\bibitem{ZW}
B. Zwiebach, ``Interpolation String Field Theories'',
{\em Mod. Phys. Lett.} {\bf A7} (1992) 1079; hep-th/9202015.

\bibitem{HN}
A. Hashimoto and N. Itzhaki, ``Observable of String Field
Theory'', JHEP {\bf 0201} (2002) 028 [arXiv:hep-th/0111092].
\bibitem{GRSW}
D. Gaiotto, L. Rastelli, A. Sen and B. Zwiebach, ``Ghost
Structure and Closed Strings in Vacuum String Field Theory'',
Adv. Theor. Math. Phys. {\bf 6} (2002) 403 [arXiv:hep-th/0111129].
\bibitem{MAMG}
M. Alishahiha and M.R. Garousi, ``Gauge Invariant Operators and
Closed String Scattering in Open String  Field Theory'', Phys.
Lett. B {\bf 536} (2002) 129 [arXiv:hep-th/0201249].
\bibitem{MGGM}
M.R. Garousi and G.R. Maktabdaran, ``Excited D-brane decay in
Cubic String Field Theory and in Bosonic String Theory'', Nucl.
Phys. B{\bf 651} (2003) 26 [arXiv:hep-th/0210139].
\bibitem{TTSZ}
T. Takahashi and S. Zeze, ``Closed Sting Amplitude In Open String
Field Theory'', JHEP {\bf 0308} (2003) 020 [arXiv:hep-th/0307173].


\bibitem{GM}
M.R. Garousi and R.C. Myers, Nucl. Phys. B{\bf 475} (1996) 193
[arXiv:hep-th/9603194].
\bibitem{HandBook}
M. Abramowitz, I.A. Stegun, ``Handbook of Mathematical Functions,
With Formulas, Graphs, and Mathematical Tables'', Dover
Publications, INC., New York (1974). \\For a  online library,
 see ``The Wolfram Function Site'' at
http://functions.wolfram.com.
\bibitem{HK}
A. Hashimoto, I.R. Klebanov, Phys. Lett. B {\bf 381} (1996) 437;
[arXiv:hep-th/9604065].



\end{thebibliography}
\end{document}